\documentclass{article}

\usepackage{arxiv}

\usepackage[utf8]{inputenc} % allow utf-8 input
\usepackage[T1]{fontenc}    % use 8-bit T1 fonts
\usepackage{hyperref}       % hyperlinks
\usepackage{url}            % simple URL typesetting
\usepackage{booktabs}       % professional-quality tables
\usepackage{amsfonts}       % blackboard math symbols
\usepackage{nicefrac}       % compact symbols for 1/2, etc.
\usepackage{microtype}      % microtypography
\usepackage{lipsum}
\usepackage{fancyhdr}       % header
\usepackage{graphicx}       % graphics
\graphicspath{{media/}}     % organize your images and other figures under media/ folder

\usepackage{lineno}
\usepackage{amsmath}
\usepackage{amssymb}
\usepackage{algorithm}
\usepackage{algpseudocode}
\usepackage{subcaption}
\usepackage{tikz}
\usepackage{booktabs}
\usepackage{multirow}
\usepackage{tabularx}

\modulolinenumbers[5]

\title{EclipseNETs: Learning Irregular Small Celestial Body Silhouettes}

\author{
  Giacomo Acciarini \\
  Advanced Concepts Team, ESA \& University of Surrey \\
  \texttt{g.acciarini@surrey.ac.uk} \\
  \And
  Dario Izzo \\
  Advanced Concepts Team, ESA\\
  \texttt{dario.izzo@esa.int}
   \And
  Francesco Biscani \\
  Skylon Dynamics \\
  \texttt{fbiscani@skylon.dev} \\
}

\begin{document}
\maketitle
\begin{abstract}
Accurately predicting eclipse events around irregular small bodies is crucial for spacecraft navigation, orbit determination, and spacecraft systems management. This paper introduces a novel approach leveraging neural implicit representations to model eclipse conditions efficiently and reliably. We propose neural network architectures that capture the complex silhouettes of asteroids and comets with high precision. Tested on four well-characterized bodies - Bennu, Itokawa, 67P/Churyumov-Gerasimenko, and Eros - our method achieves accuracy comparable to traditional ray-tracing techniques while offering orders of magnitude faster performance. Additionally, we develop an indirect learning framework that trains these models directly from sparse trajectory data using Neural Ordinary Differential Equations, removing the requirement to have prior knowledge of an accurate shape model.
This approach allows for the continuous refinement of eclipse predictions, progressively reducing errors and improving accuracy as new trajectory data is incorporated.

%By enabling more autonomous operations around irregular celestial bodies, our approach supports future scientific missions and resource utilization efforts.
\end{abstract}


%These mathematical models are utilized to forecast the orbital state vectors of resident space objects, taking into account a simplified representation of the forces influencing these satellites.

\section{Introduction}
\label{sec:introduction}
Small celestial bodies, including asteroids and comets, play a critical role in space exploration. They provide insights into the early Solar System, offer potential resources for future missions, and pose impact threats to Earth~\cite{pieters1994meteorite, sanchez2011asteroid}. As remnants of Solar System formation, they preserve primordial materials that inform our understanding of early chemical and physical conditions. Additionally, they present unique opportunities for scientific exploration, such as sample-return missions and spacecraft technology testing~\cite{fujiwara2006rubble, watanabe2017hayabusa2, cheng2018aida, michel2022esa}.
As their exploration becomes more advanced, understanding the dynamics of spacecraft as they orbit or interact with these bodies is critical, particularly given their irregular shapes, weak gravity, and complex physical properties~\cite{scheeres2012orbital}.

Precise eclipse condition estimation is vital in small-body exploration, impacting solar radiation pressure calculations, communication, and spacecraft thermal and power management~\cite{adhya2004oblate, jia2017eclipse, srivastava2015eclipse}. For objects with high area-to-mass ratios orbiting weak-gravity bodies, solar radiation pressure is a dominant perturbation, making accurate eclipse computation essential for predicting spacecraft motion~\cite{lang2022heliotropic, lang2021spacecraft, mcmahon2020dynamical}.

Eclipse computation involves determining whether an orbiting object's position lies within a celestial body's shadow~\cite{neta1998satellite}. Shadow modeling depends on the body's shape and size; simple bodies allow for approximations using cylindrical cones, while irregular bodies require high-fidelity 3D models~\cite{malyuta2021advances}. For example, the NASA OSIRIS-REx mission employed a detailed 3D model of asteroid 101955 Bennu to simulate the precise shadow region and the motion of ejected particles~\cite{hergenrother2019operational, mcmahon2020dynamical}.

Traditional eclipse modeling methods, such as the Möller–Trumbore intersection algorithm~\cite{moller2005fast}, use ray-tracing techniques to determine shadow regions. While highly accurate, these methods are computationally expensive and non-differentiable, limiting their utility in reliably estimating eclipse entrance and exit conditions~\cite{biscani2022reliable}. Recent advances in neural image processing introduce differentiable and computationally efficient alternatives using implicit neural representations, particularly Neural Radiance Fields (NeRFs). These models capture complex 3D scenes with photometric consistency~\cite{mildenhall2021nerf}. Periodic activation functions, as explored by Sitzmann et al.~\cite{sitzmann2020implicit}, enhance the ability to model high-frequency details, making them well-suited for accurately describing shadows cast by irregular celestial bodies~\cite{biscani2022reliable, acciarini2024eclipsenets}.

In this paper, we construct a database of irregular silhouettes for four small celestial bodies using their shape models. The dataset is split into training and validation sets to train neural networks, termed EclipseNETs, for reconstructing small-body silhouettes from different directions. We first compare periodic activation functions with ReLU-based architectures confirming their superior ability in capturing complex shapes. Then, we integrate these models into orbital propagators to assess their accuracy and computational efficiency relative to traditional methods. The results demonstrate that EclipseNETs can efficiently and accurately perform eclipse condition computation. Due to the differentiable nature of EclipseNETs, this approach further facilitates the use of Taylor propagators \cite{biscani2022reliable}, enhancing the reliability and accuracy of eclipse event detection during orbital propagation.

In the second part of the paper, we introduce a training technique based on Neural Ordinary Differential Equations (NeuralODEs)~\cite{chenneuralode, chen2020learning}, which enables the direct learning of silhouettes from trajectory data. This approach eliminates the need for precise prior knowledge of the irregular shapes of small bodies, which are often not accurately known before in situ exploration. By incorporating EclipseNET, the resulting spacecraft dynamics can be formulated as a NeuralODE, allowing direct optimization of the network parameters from trajectory data. This enables learning and/or refinement of the eclipse model by minimizing trajectory prediction errors, presenting a novel strategy to improve irregular eclipse descriptions through NeuralODEs.

The paper is structured as follows: Section~\ref{sec:methods} discusses methods and related work, focusing on eclipse function definitions, the irregular bodies studied, and the spacecraft dynamics. Section~\ref{sec:learning_small_body_silhouettes_w_eclipsenet} details the training process of EclipseNETs, comparing Siren-based and ReLU-based architectures. We also discuss EclipseNET integration in a Taylor-based orbital propagator, benchmarking it against traditional ray-intersection methods. Section~\ref{sec:online_learning_via_neuralodes} introduces an algorithm for direct silhouette learning from trajectory data via NeuralODEs. Finally, Section~\ref{sec:conclusions} summarizes our findings.

\section{Methods}
\label{sec:methods}
\subsection{Eclipse Function}
\label{sec:eclipse_function}

An eclipse function, $F_{\mathcal B}(x,y, \hat{\mathbf s})$, is introduced to implicitly represent the eclipsed region created by an irregular body $\mathcal B$ when illuminated by a point source at infinity in the direction $\hat{\mathbf s}$ (in the body frame). This function is defined in a reference frame lying in a plane orthogonal to $\hat{\mathbf s}$.

\begin{figure}[ht!]
  \centering
  \includegraphics[width=0.8\columnwidth]{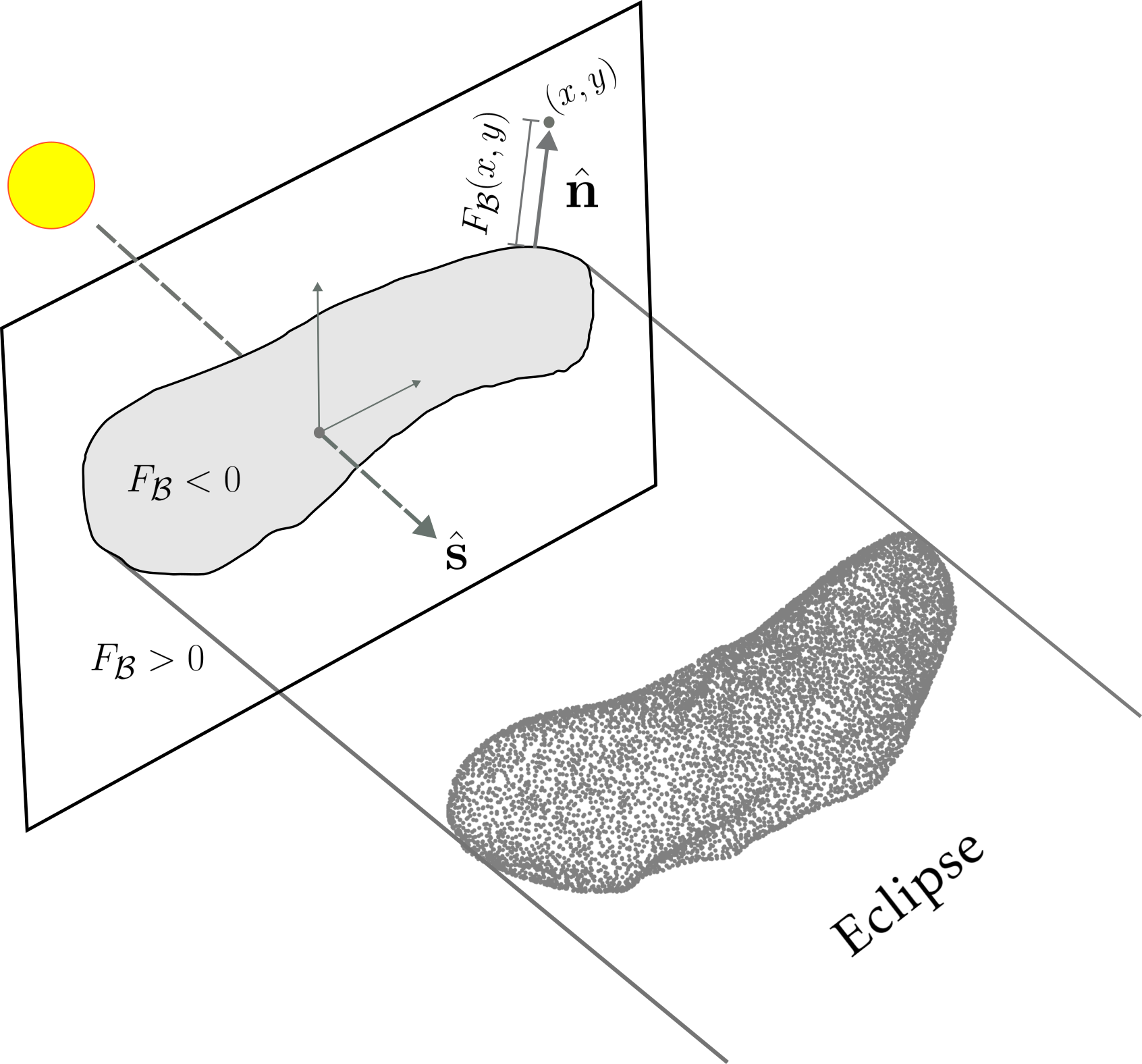}
  \caption{Schematic illustration of the shadow cast by an irregular body, and its eclipse function $F_\mathcal B$.}
	\label{fig:eclipse_geometry}
\end{figure}

Mathematically, the eclipse function is given by:
$$
F_{\mathcal B}(x,y, \hat{\mathbf s}) = \left\{
\begin{array}{l}
d(x,y, \partial \Omega_{\hat{\mathbf s}}) \ \textrm{(outside the eclipse region)}\\
-d(x,y, \partial \Omega_{\hat{\mathbf s}}) \ \textrm{(inside the eclipse region)}\\
\end{array}
\right.
$$
where $d(x,y,\partial \Omega_{\hat{\mathbf{s}}})$ represents the signed distance function from the point ($x,y$) to the boundary $\partial \Omega_{\hat{\mathbf{s}}}$ of the eclipse region. This function is closely related to the signed distance function, commonly used in computer graphics~\cite{park2019deepsdf}, but here it describes two-dimensional shapes parametrized by the projection direction $\hat{\mathbf{s}}$. The gradient of the eclipse function in the direction perpendicular to the eclipse boundary satisfies:
$$
\nabla_{x,y}F_{\mathcal{B}}\cdot \hat{\mathbf{n}}=1\text{,}
$$
where $\hat{\mathbf{n}}$ is the unit normal (either outward or inward) to $\partial \Omega_{\hat{\mathbf{s}}}$. %This extra condition can be added to the loss function during training, to help in convergence.
Figure~\ref{fig:eclipse_geometry} provides a visual representation of the eclipse function and its relevant geometric properties. While Figure~\ref{fig:plot_small_bodies} B) illustrates examples of the eclipse function for a specific direction across different bodies. 

\subsection{Irregular Bodies}
\label{sec:irregular_bodies}

Our focus is on four different small bodies: 101955 Bennu, 67P / Churyumov - Gerasimenko, 433 Eros, and 25143 Itokawa. In all cases, high-fidelity polyhedral models reconstructed from various instruments on board spacecraft that visited these celestial bodies are utilized. For Eros and Itokawa, models produced by Robert Gaskell~\cite{erospoly,itokawapoly} are used; for Bennu, the model provided by the OSIRIS-REx team~\cite{bennupoly} is employed; and for Churyumov–Gerasimenko, the model available from the European Space Agency~\cite{67ppoly} is used.

\begin{figure}[ht!]
  \centering
  \includegraphics[width=1.\columnwidth]{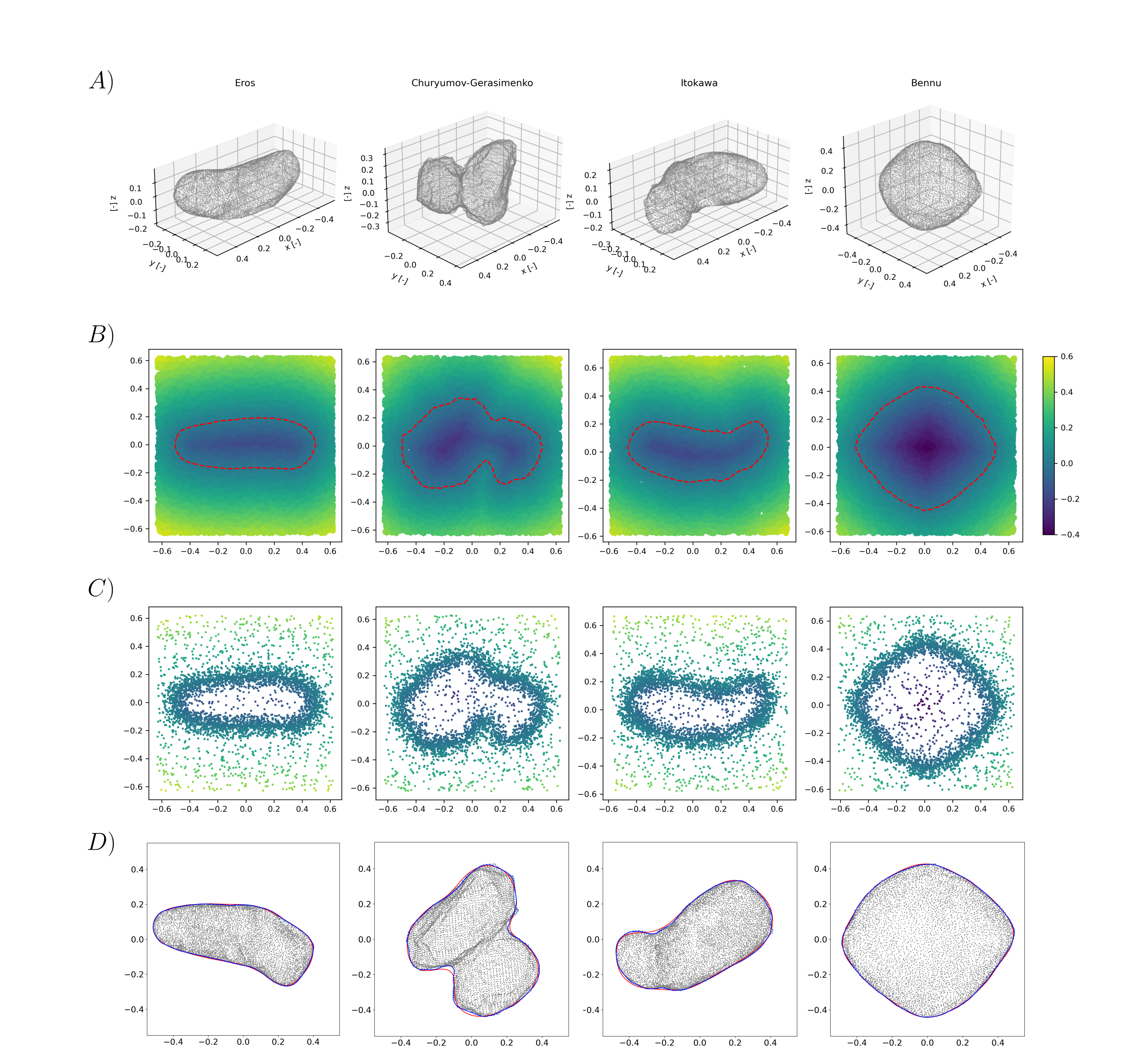}
  \caption{A) 3D models of Bennu, Churyumov-Gerasimenko, Eros, and Itokawa. B) Contour plot of the eclipse function for a fixed viewpoint. C) Sampled points of the eclipse function used to build the training set. D) Eclipse predictions for a Sun direction not included in the training set. The red curve represents an EclipseNet with 2,369 parameters, while the blue curve corresponds to 50,561 parameters.}
	\label{fig:plot_small_bodies}
\end{figure}

To model the gravitational field of these bodies, the surface meshes are transformed into mascon models following the same procedure outlined in~\cite{izzo2022geodesy}. This is achieved by first generating a constrained Delaunay tetrahedralization~\cite{si2015tetgen} of the polyhedral mesh and then assigning a mass at the centroid of each resulting tetrahedron. Figure~\ref{fig:plot_small_bodies} A) presents a 3D visualization of these models. The final representation of each asteroid consists of a triangular mesh that defines the surface geometry and a mascon model that describes the mass distribution, which is assumed to be uniform in all cases.

The resolution of these models, along with physical properties such as the rotation period and mass of each body, affects the dynamics of an orbiting spacecraft. The number of mesh points and mascons determines the fidelity of both the shape and gravitational field representation, while the rotation rate ($\omega$) influences orbital stability and eclipse conditions. A characteristic length scale ($L$) is used to normalize positional coordinates. Table~\ref{tab:parameters_asteroids_comets} summarizes these key parameters for each of the four celestial bodies considered in this study.

\begin{table}[ht!]
\centering
\begin{tabular}{|l|c|c|c|c|c|c|}
\hline
        \ & N mesh & N mascons & $\omega$ [hr] & L [km] & mass [kg] \\
\hline
        Bennu                   &  7,374                  & 75,150                 & 4.296        &  0.5634   & $7.329\times 10^{10}$ \\
        Itokawa                 & 3000                   & 100,363                 & 12.132        &  0.5607    &  $3.51\times 10^{10}$ \\
        67P                    & 9,149                   & 57,259                 & 12.4043             &  5.0025   &  $9.982\times10^{12}$  \\
        Eros                    & 7,374                   & 97,824                 & 5.270              & 32.6622   & $6.687 \times 10^{15}$  \\
\hline
\end{tabular}
\caption{Key parameters of the 3D models used for small-body analysis.}
\label{tab:parameters_asteroids_comets}
\end{table}

The 3D triangular mesh models are used to determine the ground truth eclipse conditions by applying the Möller–Trumbore intersection algorithm~\cite{moller2005fast}, which checks whether a ray from the light source intersects the body's surface, providing a ground truth for eclipse detection. Meanwhile, the mascon model is incorporated into trajectory simulations to account for the irregular gravitational field of each asteroid or comet. By combining these high-resolution shape models with a mascon distribution, we ensure an accurate and high-fidelity representation of the dynamical environment surrounding these small celestial bodies.

\subsection{Spacecraft Dynamics}
\label{sec:spacecraft_dynamics}
The spacecraft dynamics, taken from \cite{biscani2022reliable}, is described in a body-centered reference frame by formulating the following set of differential equations:
\begin{equation}
\begin{cases}
    \dot{\pmb{r}}&=\pmb{v}\\
    \dot{\pmb{v}}&=-G \displaystyle\sum_{j=0}^{N} \dfrac{m_j}{|\pmb{r} - \pmb{r}_j|^3} (\pmb{r} - \pmb{r}_j) - 2 \boldsymbol{\omega} \times \pmb{v} - \boldsymbol{\omega} \times (\boldsymbol{\omega} \times \pmb{r}) - \eta \nu(\pmb{r})\hat{\mathbf{s}}(t)
\text{,}
\end{cases}
\label{eq:spacecraft_dynamics}
\end{equation}
where $\pmb{r}$, $\pmb{v}$ denote the position and velocity vectors, respectively, of the spacecraft in the body-centered reference frame, $m_j$ and $\pmb{r}_j$ represent the mass and position, respectively, of each mascon that characterizes the asteroid’s irregular gravitational field, and $\boldsymbol{\omega}$ is the angular velocity of the asteroid’s rotation. 
The term $\eta$ corresponds to the acceleration magnitude due to solar radiation pressure, which is modeled to be $10^{-3}$ m/s$^2$ for all simulations, while the term $\nu(\mathbf{r})$ represents the eclipse factor, which is considered as zero ($\nu=0$) when the spacecraft is in eclipse and one ($\nu=1$) otherwise.
Traditional implementations use ray-tracing algorithms to determine the eclipse factor and to discontinuously change the dynamics. In this case, however, the aim is to substitute these algorithms with neural models. Hence, as detailed in Section~\ref{sec:learning_small_body_silhouettes_w_eclipsenet}, a small neural network, called EclipseNET, will be used to represent the eclipse factor.
Then, $\boldsymbol{\omega} \times \mathbf{v}$, represents the Coriolis force, resulting from the rotational motion of the asteroid, and $\boldsymbol{\omega} \times (\boldsymbol{\omega} \times \mathbf{r})$ corresponds to the centrifugal force experienced due to the asteroid's rotation. 
Finally, the unit vector, $\hat{\mathbf{s}}(t)$, describes the direction of the Sun. 
The Sun direction is modeled as \(\hat{\mathbf{s}}(t) = \mathbf{R}(t) \hat{\mathbf{s}}(0)\), where \(\mathbf{R}(t)\) is the rotation matrix that evolves over time as the asteroid rotates at an angular velocity \(\boldsymbol{\omega}\), causing a time-dependent change in the Sun's direction relative to the spacecraft. 
Using this model, we consider the Sun's direction as a simple rotation of the initial direction \(\hat{\mathbf{s}}(0)\) by an angle \(\omega t\).
All these effects combine to determine the spacecraft's trajectory relative to the rotating asteroid reference frame.

\section{Learning Small Body Silhouettes with EclipseNET}
\label{sec:learning_small_body_silhouettes_w_eclipsenet}

To model the complex eclipse conditions caused by irregular small bodies, we introduce EclipseNET, a neural network architecture designed to implicitly represent the geometry of eclipses. Given the recent success of Siren networks~\cite{sitzmann2020implicit} in neural implicit representations of complex shapes, we compare EclipseNETs trained with ReLU and with periodic activation functions. 

We write the EclipseNET as:
\begin{equation}
    e_{\pmb{\theta}}=\mathcal{N}_{\pmb{\theta}}(\pmb{r},\hat{\mathbf{s}})
\end{equation}
where $\pmb{r}$ is the positional vector of the spacecraft, and $\hat{\mathbf{s}}$ is the Sun direction. 
To facilitate the training, the state is projected onto the plane perpendicular to the Sun direction, and the two direction components of the projected spacecraft position are normalized and used as inputs, while the Sun direction is encoded using the sine and cosine of azimuth and elevation angles, making the total size of the network inputs six-dimensional. The output, $e_{\pmb{\theta}}$, is a scalar value representing the eclipse function: negative when the spacecraft is in eclipse, positive when it is outside of eclipse, and zero at the asteroid silhouette.

To train EclipseNET, we create a dataset that contains ground truth values for the eclipse function found using the shape models of the small bodies. We consider an isotropic range of Sun directions, which is generated using the Fibonacci sphere. This results in 500 Sun directions for training and 200 for validation. For each direction, we uniformly sample 1,000 points $(x, y)$ in the range $[-1, 1]$ and 3,000 points around the eclipse boundary $\partial \Omega_{\hat{\mathbf{s}}}$. An example of the result of this sampling procedure for the four small bodies in a random direction is shown in Figure~\ref{fig:plot_small_bodies} C). This data yields a total of 22 million training points and 9 million validation points for each of the four small bodies studied. 

\begin{table}[ht!]
\scriptsize
    \centering
    \begin{tabular}{|l|c|c|c|c|c|c|}
\hline
        \ &  $\mathcal{L}_{MSE}$ ReLU &  $\mathcal{L}_{MSE}$ Siren &  Dataset\\
\hline
         $\mathcal N_{Bennu}$ & $3.5295\times 10^{-5}$ &$1.9906\times 10^{-5}$ & $\mathcal{D}_{train,Bennu}$\\
         $\mathcal{N}_{Itokawa}$  &  $8.0184\times10^{-5}$ &  $4.7082\times 10^{-5}$ & $\mathcal{D}_{train,Itokawa}$\\
       $\mathcal{N}_{67P}$ & $7.9564\times 10^{-5}$  &    $4.6469\times 10^{-5}$ &  $\mathcal{D}_{train,67P}$\\
        $\mathcal{N}_{Eros}$ &  $9.3738\times 10^{-5}$ &   $4.2137\times 10^{-5}$ &  $\mathcal{D}_{train,Eros}$\\
\hline
         $\mathcal N_{Bennu}$ &  $3.6508\times 10^{-5}$ & $2.1773\times 10^{-5}$ & $\mathcal{D}_{valid,Bennu}$\\
         $\mathcal{N}_{Itokawa}$  & $8.4136\times 10^{-5}$ & $4.9863\times 10^{-5}$ & $\mathcal{D}_{valid,Itokawa}$\\
       $\mathcal{N}_{67P}$ & $8.1466\times 10^{-5}$ &  $4.9408\times 10^{-5}$ &  $\mathcal{D}_{valid,67P}$\\
        $\mathcal{N}_{Eros}$ & $9.8826\times 10^{-5}$ & $4.5850\times 10^{-5}$ &  $\mathcal{D}_{valid,Eros}$\\
    \hline
    \end{tabular}
    \caption{Feed forward neural networks with ReLU activation function and Siren network models mean squared error loss (MSE) on the training and validation datasets for all four small bodies.}
    \label{tab:ml_models_results}
\end{table}

We compare two types of neural architectures for training EclipseNETs: a Siren architecture and a more traditional ReLU-based neural network, both with 2,369 learnable parameters. As reported in Table~\ref{tab:ml_models_results}, we observe that Siren networks consistently outperform the ReLU-based networks both on the training and validation sets. The Siren network achieves a lower mean squared error (MSE), which highlights its superior ability to capture the subtle details of the eclipse boundary. 
\begin{figure}[ht!]
  \centering
  \includegraphics[width=0.8\columnwidth]{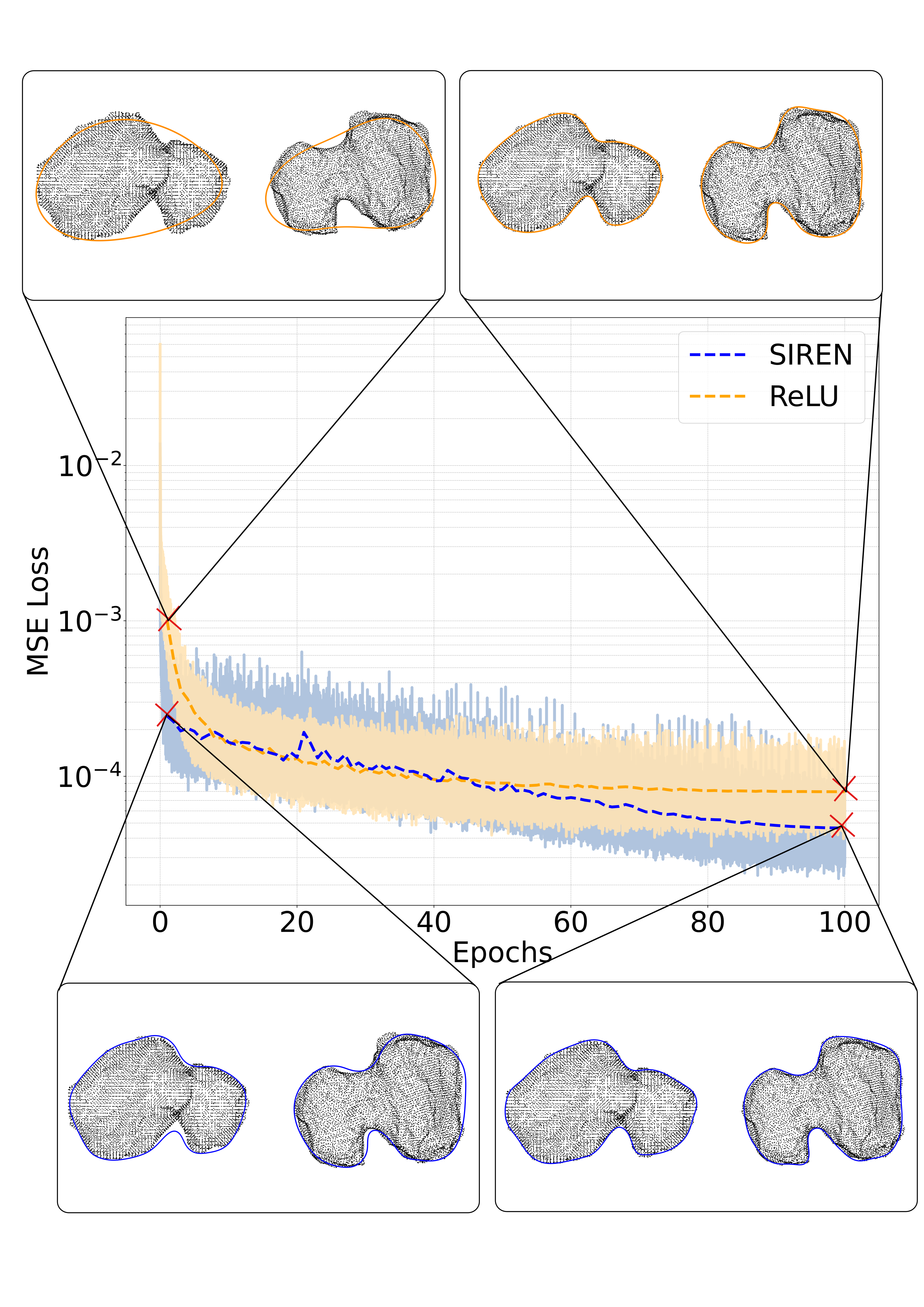}
  \caption{Training loss vs the number of epochs, together with some views of the approximated silhouette for both networks at the first and last epochs, for both Siren and ReLU networks, for the case of 67P/Churyumov–Gerasimenko.}
	\label{fig:eclipsenet_training}
\end{figure}
To support this, Figure~\ref{fig:eclipsenet_training} presents the MSE loss as a function of training epochs for both the Siren and ReLU networks, for the 67P case. The figure includes two unseen views during training of the asteroid from the first and last epochs, along with their corresponding silhouette approximations, generated using the ReLU network (top, in orange) and the Siren network (bottom, in blue). As observed, the Siren network not only achieves lower loss levels than the ReLU network but also converges more quickly. This results in a more accurate silhouette reconstruction, particularly at the initial epoch, where the difference is noticeable. By the final epoch, the ReLU network reduces the gap to some extent but still struggles with sharp and irregular features in the shape.

While inspecting these losses and visually observing the accuracy of the silhouette reconstruction provides a good indication of the accuracy of the EclipseNET, it is however equally important to quantify what the resulting error in the state of a spacecraft orbiting these small bodies is, shall the EclipseNET be used to identify eclipse regions. Formally, this means that when integrating Eq.~\eqref{eq:spacecraft_dynamics}, the eclipse factor is directly determined using the EclipseNET as an event function in the integrator. In this way, when the EclipseNET is zero, then the spacecraft is crossing the asteroid silhouette, which causes the eclipse factor to change value (from zero to one or vice-versa) if the spacecraft is behind the small body with respect to the Sun.  
\begin{figure}[ht!]
  \centering
  \includegraphics[width=1.\columnwidth]{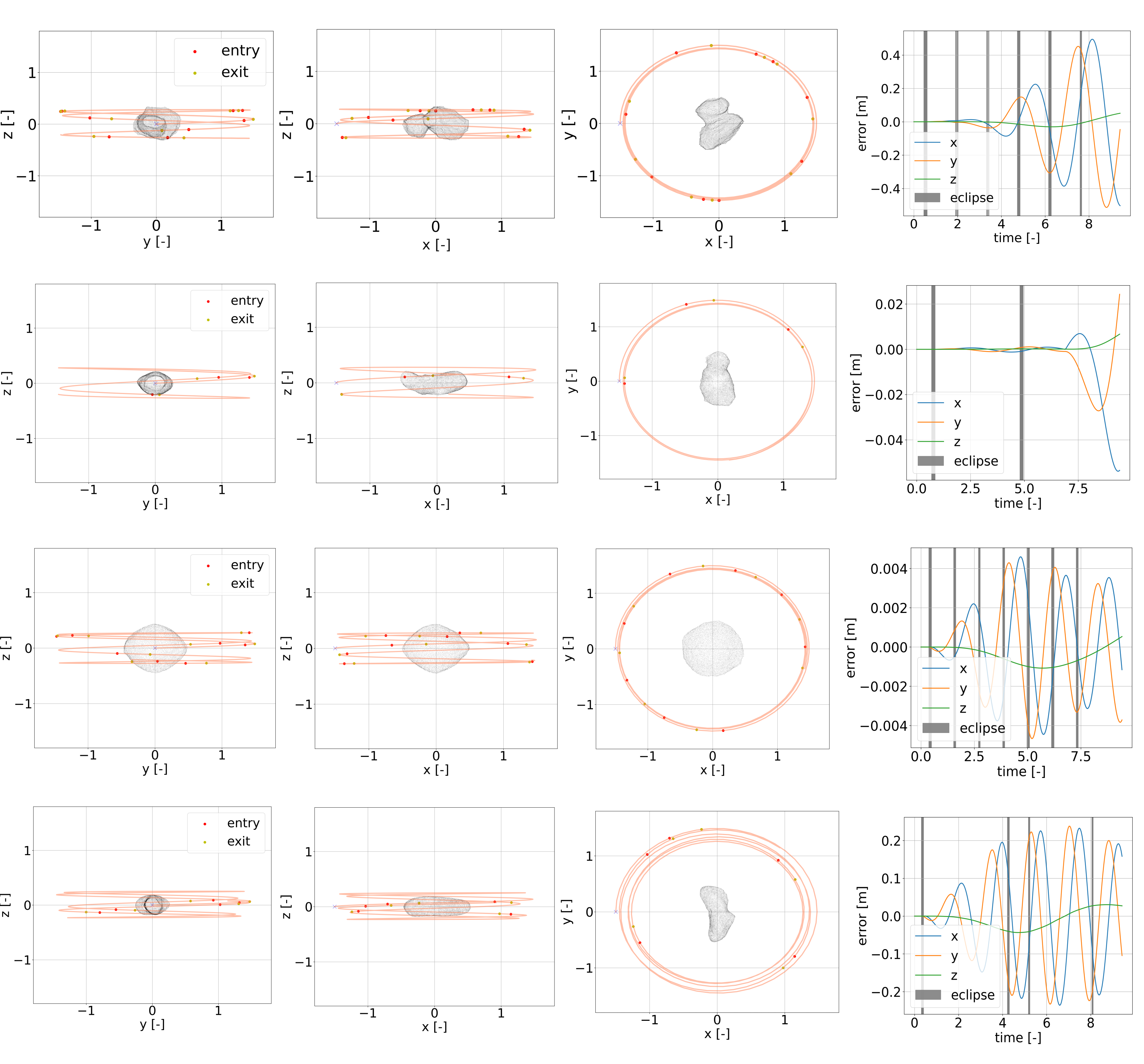}
  \caption{\textit{left three columns}: three different projections of the three-dimensional orbits obtained using EclipseNET as an event; \textit{right column}: positional coordinate errors between the trajectory computed with EclipseNET as an event and the one obtained using the Möller-Trumbore algorithm.}\label{fig:plot_orbits_w_eclipses}
\end{figure}

In the first three columns of Figure~\ref{fig:plot_orbits_w_eclipses}, we present three projections of three-dimensional orbits around the four tested small bodies, highlighting the entry and exit points of eclipses along the trajectory. For these experiments, the Siren networks were used as EclipseNETs. As shown in the rightmost column, using these models to approximate the asteroid silhouette yields eclipse prediction errors within centimeters for all tested small bodies, even after multiple orbits. This accuracy is benchmarked against the eclipses computed using a ray-tracing method based on the Möller–Trumbore triangle intersection algorithm~\cite{moller2005fast}. The computational efficiency of EclipseNET is remarkable, with inference speeds more than two orders of magnitude faster than the ray-tracing approach, demonstrating its potential to replace computationally expensive traditional algorithms in space-flight simulations.

While these results are promising, there will always be a slight discrepancy between the actual eclipse boundary and its representation by EclipseNET. This mismatch is expected to grow over time, particularly when the spacecraft traverses more complex eclipse regions or encounters irregular bodies with intricate geometries. Additionally, the irregular shape of small bodies is often not accurately known beforehand, and precise information about the body’s shape may only become available as the spacecraft closely approaches the asteroid or comet.

Hence, there might be the need for continuous refinement and adaptability of the model's accuracy, as trajectory data become available. To address this, the next section introduces online training methods based on NeuralODEs~\cite{chenneuralode}. These techniques allow the model to dynamically adjust in real-time, based on the evolving spacecraft trajectory, ensuring that the model becomes increasingly precise as it learns from positional data during the mission. 
Building upon this, the following section introduces how NeuralODEs can be employed to learn the silhouette of an irregular small body directly from trajectory data. 

\section{Learning the Silhouette from Trajectories via NeuralODEs}
\label{sec:online_learning_via_neuralodes}
As discussed in the previous section, the shape of an irregular small body is often incompletely known or lacks sufficient resolution prior to a spacecraft visit. Even when a preliminary shape model is available, EclipseNET’s implicit representation may contain inaccuracies that could lead to growing errors over time. Consequently, further model refinement is often necessary to ensure an acceptable error in predicting the satellite's position and velocity.

To address this challenge, we propose a method to dynamically learn and update the neural representation of the small body’s silhouette directly from sparse trajectory data. 
This approach leverages NeuralODEs~\cite{chenneuralode,chen2019neural}, specifically ODEs with neural event functions~\cite{chen2020learning}, to iteratively refine the eclipse function. 
This enables real-time and continuous adaptation of the model as new observational data becomes available.

As clarified in Section~\ref{sec:spacecraft_dynamics}, the spacecraft dynamics changes according to a termination criterion triggered by an eclipse function, which in our case is modeled as a neural network (EclipseNET). 
Our objective is to refine the EclipseNET used as a neural event function to reduce the observed errors in the spacecraft’s states. To achieve this, we first define the augmented system dynamics, where the first-order ODE governing the satellite's motion in the small-body environment is extended with an equation for the event manifold time derivative:

\begin{equation}
    \begin{cases}
        \dot{\pmb{x}}&=\pmb{f}_{\pmb{\theta}}(t,\pmb{x})\\
        \dot{\varepsilon}_{\pmb{\theta}}&=g_{\pmb{\theta}}(t,\pmb{x})
    \text{,}
    \end{cases}
    \label{eq:augmented_dynamics}
\end{equation}
where the state is composed of the position and velocity of the spacecraft, $\pmb{x}=[\pmb{r},\pmb{v}]$ and the variable $\varepsilon_{\pmb{\theta}}$ has been introduced to track the time derivative of the EclipseNET. The term $g_{\pmb{\theta}}$ is thus derived as: 
$$
g_{\pmb{\theta}}=\nabla_{\hat{\mathbf{s}}}e_{\pmb{\theta}}\cdot \dot{\hat{\mathbf{s}}}(t)+\nabla_{\pmb{x}}e_{\pmb{\theta}}\cdot \pmb{f}_{\pmb{\theta}}\text{.}
$$ 
Under our assumptions, the derivative of the Sun direction can be found as: $\dot{\hat{\mathbf{s}}}=\dot{\mathbf{R}}(t)\hat{\mathbf{s}}(0)$.
Hence, the first equation mirrors Eq.~\eqref{eq:spacecraft_dynamics}, while the EclipseNET, $e_{\pmb{\theta}}$, acts as the event trigger function, to change the value of $\nu(\pmb{r})$. The learnable parameters of the EclipseNET, $\pmb{\theta}\in \mathbb{R}^m$, control the discontinuous transitions in the dynamics, with the subscript indicating this dependency. The result is a NeuralODE, where the neural network operates as an event function.  

Then, the loss function is defined as the mean squared error (MSE) between the modeled states $\pmb{x}_{\pmb{\theta}}$ (computed integrating the dynamics with the neural event function) and the observed ones $\overline{\pmb{x}}$:
\begin{equation}
\mathcal{L}=\sum_{j=0}^T||\pmb{x}_{j,\pmb{\theta}}-\pmb{\overline{x}}_{j}||^2\text{,}
    \label{eq:loss}
\end{equation}
where $j$ indexes the time instances $t_j$ at which observations are available. The observed states may originate from actual trajectory measurements or high-fidelity simulations, such as those employing the Möller-Trumbore intersection algorithm.

The parameters of EclipseNET are updated by computing the gradient of the loss with respect to $\pmb{\theta}$:
\begin{equation}
    \nabla_{\pmb{\theta}} \mathcal{L}=\sum_{i=1}^n\dfrac{\partial \mathcal{L}}{\partial x_{i}}\dfrac{\partial x_i}{\partial e_{\pmb{\theta}}}\nabla_{\pmb{\theta}}e_{\pmb{\theta}} \text{.}
\end{equation}
The first term follows directly from Eq.~\eqref{eq:loss}:
\begin{equation}
    \dfrac{\partial \mathcal{L}}{\partial x_i}=2\sum_{j=1}^{T}(x_{ij,\pmb{\theta}}-\overline{x}_{ij})\text{.}
\end{equation}
The second term can be derived symbolically by first computing $\dfrac{\partial e_{\pmb{\theta}}}{\partial x_i}$, while the third term is obtained by integrating Eq.~\eqref{eq:augmented_dynamics} along with its variational equations with respect to the parameters:
\begin{equation}
\begin{cases}
    \dot{\pmb{x}}&= \pmb{f}_{\pmb{\theta}}(t,\pmb{x})\\
    \dot{\varepsilon}_{\pmb{\theta}}&=g_{\pmb{\theta}}(t,\pmb{x})\\
    \dfrac{d}{dt}\bigg( \dfrac{\partial x_{i}}{\partial \theta_k}\bigg) &= \displaystyle\sum_{j=1}^n\dfrac{\partial f_{i,\pmb{\theta}}}{\partial x_j}\dfrac{\partial x_j}{\partial \theta_k}+\dfrac{\partial f_{i,\pmb{\theta}}}{\partial \theta_k}\\
    \dfrac{d}{dt}\bigg( \dfrac{\partial e_{\pmb{\theta}}}{\partial \theta_k} \bigg)&= \displaystyle\sum_{j=1}^n\dfrac{\partial g_{\pmb{\theta}}}{\partial x_j}\dfrac{\partial x_j}{\partial \theta_k}+\dfrac{\partial g_{\pmb{\theta}}}{\partial \theta_k}\text{,}
\end{cases}
\end{equation}
$\forall k=1,\dots,m$.
This results in a system of $(n+1) + (n+1)m$ equations to be integrated: we use the open-source \textit{heyoka} software to both construct the variational equations and integrate them via Taylor method~\cite{biscani_heyoka}.
\begin{figure}[ht!]
  \centering
  \includegraphics[width=1.\columnwidth]{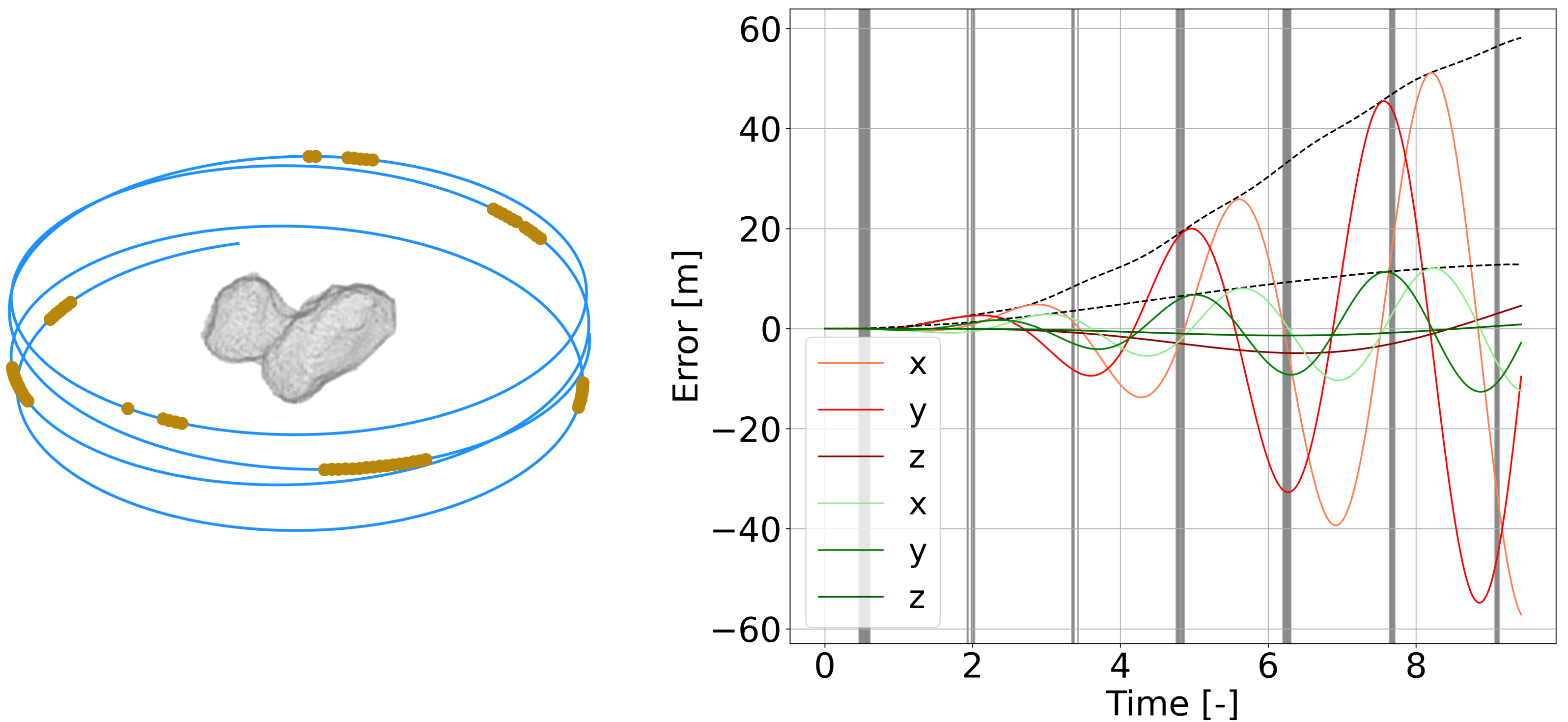}
  \caption{\textit{left}: orbit around 67P/ Churyumov-Gerasimenko, with eclipse regions highlighted in dark yellow along the trajectory; \textit{right}: error in the three positional components before (in red) and after (in green) the NeuralODE refinement. Eclipses are here displayed in grey.}
	\label{fig:neural_ode_w_event_refinement}
\end{figure}

In the right panel of Figure~\ref{fig:neural_ode_w_event_refinement}, we show the errors in the $x$, $y$, and $z$ positional components both before (in red) and after (in green) the NeuralODE refinement. The dashed black line in the same plot represents the norm of the error difference for both cases. The loss was computed using a single observation at the final time step of the simulation ($t_{obs} = 9.425$). The target state used for comparison was obtained using the Möller-Trumbore intersection algorithm: we depict the full orbit with the eclipses highlighted in dark yellow on the left side of Figure~\ref{fig:neural_ode_w_event_refinement}.

The initial EclipseNET model was taken from the first epoch of the ReLU network discussed in Section~\ref{sec:learning_small_body_silhouettes_w_eclipsenet} and shown in Figure~\ref{fig:eclipsenet_training}. Subsequently, the NeuralODE approach was used to minimize the error between the state predicted by the Möller-Trumbore algorithm and the state obtained using EclipseNET. This process resulted in a reduction of the error norm by nearly a factor of 7.
As additional observations from diverse geometries are gathered, the representation of the small body's irregular silhouette progressively refines, enhancing the accuracy of trajectory predictions over time.

\section{Conclusions}
\label{sec:conclusions}

In this paper, we presented a novel approach for modeling eclipse conditions around irregular small celestial bodies using neural implicit representations and we also discussed how to refine these neural models directly leveraging trajectory data via NeuralODEs. 

Firstly, we introduced EclipseNET, a neural network architecture capable of accurately representing the complex silhouettes of irregular asteroids and comets. Through extensive testing on four well-characterized small bodies (Bennu, Itokawa, 67P/Churyumov-Gerasimenko, and Eros), we demonstrated that EclipseNET outperforms traditional ray-tracing methods in both accuracy and computational efficiency. In terms of architecture, the Siren-based ones achieved superior performance compared to ReLU-based networks in capturing the intricate details of eclipse boundaries. 

Secondly, we integrated EclipseNET into Taylor-based orbital propagators, enabling precise eclipse event detection during spacecraft trajectory simulations. To evaluate EclipseNET accuracy, we benchmarked it against the Möller - Trumbore ray-tracing algorithm for spacecraft position prediction across multiple orbits for all four small bodies. The results demonstrated that EclipseNET achieved centimeter-level positional accuracy while delivering inference speeds over two orders of magnitude faster than conventional methods.

Thirdly, we tackled the challenge of incomplete or imprecise prior knowledge of small body shapes by developing an online learning framework based on NeuralODEs. This approach enables continuous refinement of the eclipse model using spacecraft trajectory data, significantly reducing errors over time as new observations become available, without requiring an accurate shape model.

Our method enhances the accuracy of eclipse predictions while also supporting real-time adaptability, which is beneficial for autonomous spacecraft operations. Additionally, its differentiable nature allows for integration with modern optimization and control techniques, improving navigation around irregular celestial bodies.
\bibliography{references}

\end{document}